\newcommand{\be}{\begin{eqnarray}}
\newcommand{\ee}{\end{eqnarray}}

\documentclass[twocolumn,aps,showpacs]{revtex4}
\usepackage{graphics}
\begin{document}
%\tightenlines
%\draft
\title{Mass of the $\rho^0$  meson in ultra-relativistic heavy-ion
       collisions}  
\author{Alejandro Ayala}  
\affiliation{Instituto de Ciencias Nucleares, Universidad Nacional 
         Aut\'onoma de M\'exico, Apartado Postal 70-543, 
         M\'exico Distrito Federal 04510, M\'exico.}
\author{Jes\'us Guillermo Contreras}
\affiliation{Departamento de F\1sica Aplicada, Centro de
  Investigaci\'on y Estudios Avanzados, Unidad M\'erida, Apartado Postal
  73 Cordemex, M\'erida Yucat\'an 97310, M\'exico.}
\author{J. Magnin}
\affiliation{Centro Brasileiro de Pesquisas Fisicas, Rua Dr. Xavier
             Sigaud 150 - URCA CEP 22290-180 Rio de Janeiro Brazil.}
\begin{abstract}

We study the behavior of the $\rho$ vector mass in the context of
the almost baryon-free environment of an ultra-relativistic
heavy-ion collision. We show that $\rho$ scattering within the
hadronic phase of the collision leads to a temperature dependent,
decrease of its intrinsic mass at rest, compared to the value in
vacuum. The main contributions arise from $s$-channel scattering with
pions through the formation of $a_1$ resonances as well as with
nucleons through the formation of even parity, spin 3/2 [N(1720)] and
5/2 [$\Delta$(1905)] nucleon resonances. We show that it is possible
to achieve a shift in the intrinsic $\rho^0$ mass of order $\sim - 40$
MeV, when including the contributions of all the relevant mesons and
baryons that take part in the scattering, for temperatures
between chemical and kinetic freeze-out. 

\end{abstract}

\pacs{25.75.-q, 11.10.Wx, 11.55.Fv}

\maketitle

\section{Introduction}

In-medium modifications to the $\rho$ vector meson properties have
long been sought after as a probe of the changes experienced by
strongly interacting matter with increasing density and/or temperature
in collisions of heavy-ions at high energies. The special role played
by this meson is due to its short lifetime ($\tau\sim 1.3$fm) compared
to the lifetime of the system ($\sim 10$fm) formed in the collision. 

One possible approach is to study the $\rho$ electromagnetic decay
channels. These have the advantage of allowing the reconstruction of
the $\rho$ spectrum by looking at particles with a small interaction
probability with the surrounding hadronic medium. In this context, the
low-mass dilepton spectra have been intensively studied in experiments
from BEVALAC/SIS to SPS energies~\cite{experiments}. Nonetheless, in
spite of the success of various models~\cite{Models} that are able
to reproduce the main features of the spectra, it is fair to say that
the experimental results are thus far inconclusive with regards to the
strength of the modifications of 
the $\rho$ mass and width. Part of the problem with this approach is
that, by looking at dileptons that are continuously produced during
the different stages of the reaction, one is in fact looking at the
changes in the properties of the $\rho$ meson in a time integrated
manner, lasting over the whole evolution of the system, which makes it
difficult to distinguish the origin and magnitude of such changes.

An alternative approach is to study the $\rho$ hadronic decay
channels during the last stage of the collision, namely, during
kinetic freeze-out. This kind of probe permits us to look at the
decay, regeneration and re-scattering of the meson within a dilute
enough hadronic system over a short interval of time,  of the order or
slightly larger than the lifetime of the meson. Recently, the 
advent of large multiplicity events at RHIC energies has made it
possible to undertake such measurements. In fact, the STAR
collaboration has reported a shift of $\sim -40$ MeV and
$\sim -70$ MeV for the peak of the invariant mass distribution of the
decay $\rho^0\rightarrow\pi^+\pi^-$ in minimum bias p + p and peripheral
Au + Au collisions, respectively, at $\sqrt{s_{NN}}=200$ GeV as
compared to the vacuum value~\cite{STAR}.  

Given the resonance nature of the $\rho$ meson, changes in the
properties of the distribution of its decay products within a
thermalized medium with respect to vacuum can be generically divided
into phase space distortions of the decay products and intrinsic
changes in the properties (mass and width) of the resonance in the
heat bath~\cite{Kolb, Shuryak}. The quantitative description of the
latter is inevitably linked to model dependent considerations. Such
models are built to represent the interactions of $\rho$ mesons with
other mesons and with baryons. Based on general grounds, these models
are made to respect the basic symmetries of the strong interaction,
among them, current conservation and parity invariance. Thermal 
modifications to the $\rho$ intrinsic properties are computed by
evaluating the one-loop modification of its self-energy. Attention is 
paid to those hadrons whose rest mass is near the threshold for $s$
or $t$ channel resonance formation and with a sizable coupling to
$\rho$ and to the most abundant particles in the hadronic phase of the
collision, namely pions/kaons and nucleons. The phenomenological
coupling constants are evaluated by comparing the model prediction of
the vacuum decay rate, into the given channel, with the experimentally
measured branching ratio.

%%%%%%%%%%%%%%%%%%%%%%%%%%%%%%%%%%%%%%%%%%%%%%%%%%%%%%%%%
\begin{table}[t!]
\vspace{0.4cm}
\begin{tabular}{|cccccc|}  \hline\hline
\multicolumn{1}{|c}{$R$}  &
\multicolumn{1}{c}{$J^P$} &
\multicolumn{1}{c}{$\Gamma^{\mbox{\tiny{vac}}}_{\rho N} 
                   {\mbox{(MeV)}}$} &
\multicolumn{1}{c}{$\Gamma^{\mbox{\tiny{vac}}}_{\mbox{\tiny{tot}}} 
                   {\mbox{(MeV)}}$} &
\multicolumn{1}{c}{$f_{J^P}$} &
\multicolumn{1}{c|}{$f_{J^P}^{\mbox{\tiny{N--R}}}$} \\ \hline 
$N$(1520)      & $3/2^-$ & $25 $ & 120 &  $9.7$ &   $7$   \\   
$N$(1720)      & $3/2^+$ & $100$ & 150 &   $7$  & $7.8$   \\   
$\Delta$(1700) & $3/2^-$ & $120$ & 300 & $4.4$  &  $5$    \\
$\Delta$(1905) & $5/2^+$ & $210$ & 350 & $12.9$ & $12.2$  \\ \hline
\end{tabular}
\label{tab1}
\caption{List of baryon resonances included in the calculation. The
  last column corresponds to the values of the coupling
  constants for a non-relativistic (N-R) calculation as computed in
  Ref.~\cite{Peters}.}  
\end{table}
%%%%%%%%%%%%%%%%%%%%%%%%%%%%%%%%%%%%%%%%%%%%%%%%%%%%%%%%%%

Recall that the self-energy $\Pi$
is related to the intrinsic $\rho$ properties by
${\mbox{Im}}\ \Pi = -M_\rho\Gamma^{\mbox{\tiny{tot}}}$,
$M_\rho =\sqrt{m_\rho^2+{\mbox{Re}}\ \Pi}$,
where $m_\rho$ is the mass of $\rho$ in vacuum, $M_\rho$ and 
$\Gamma^{\mbox{\tiny{tot}}}$ are the (temperature and/or density
dependent) intrinsic mass and total decay width
of the $\rho$ meson, respectively.

Although temperature driven modifications to intrinsic
properties of $\rho$ have been thoroughly worked out from interactions
with mesons~\cite{Rapp}, the case of interactions with baryons has
mainly been given attention by looking at changes caused by dense
nuclear matter effects (see however Refs.~\cite{Rapp3}) and by means
of non-relativistic approximations for the interaction
Lagrangians~\cite{Teodorescu, Friman, Peters}.

In this work, we compute the changes to the intrinsic mass of the
$\rho$ meson as a function of temperature by considering its
scattering off pions/kaons and nucleons in a thermalized hadronic
medium, such as the one that is expected to be produced during the
dilute, almost baryon free, last stage of an ultra-relativistic
heavy-ion collision. To describe all the interactions of $\rho$
we use a manifestly covariant formalism. We show 
that there is no need to invoke a drop in the nucleon mass to
account for a sizable shift in the intrinsic mass of $\rho$. We work
in the imaginary-time formulation of thermal 
field theory to compute the one-loop $\rho$ self-energy
$\Pi^{\mu\nu}$, when interacting with the relevant hadrons. For
definitiveness, we take the $\hat{z}$ axis as the direction of motion
of the $\rho$ meson and thus the square of its thermal mass can be
computed from the thermal part of the component $\Pi^{11}$ of the
$\rho$ self-energy, in the limit of vanishing
three-momentum~\cite{Gale}.  In this 
letter, we present only the main lines of thought and central results 
reserving the details of the calculation to be reported elsewhere.

\section{Interactions of $\rho$ with nucleons and baryon
  resonances}\label{secII} 

%%%%%%%%%%%%%%%%%%%%%%%%%%%%%%%%%%%%%%%%%%%%%%%%%%%%%%%%
\vspace{0.9cm}
\begin{figure}[th] % fig1
\vspace{0.4cm}
{\centering
\resizebox*{0.4\textwidth}
{0.2\textheight}{\includegraphics{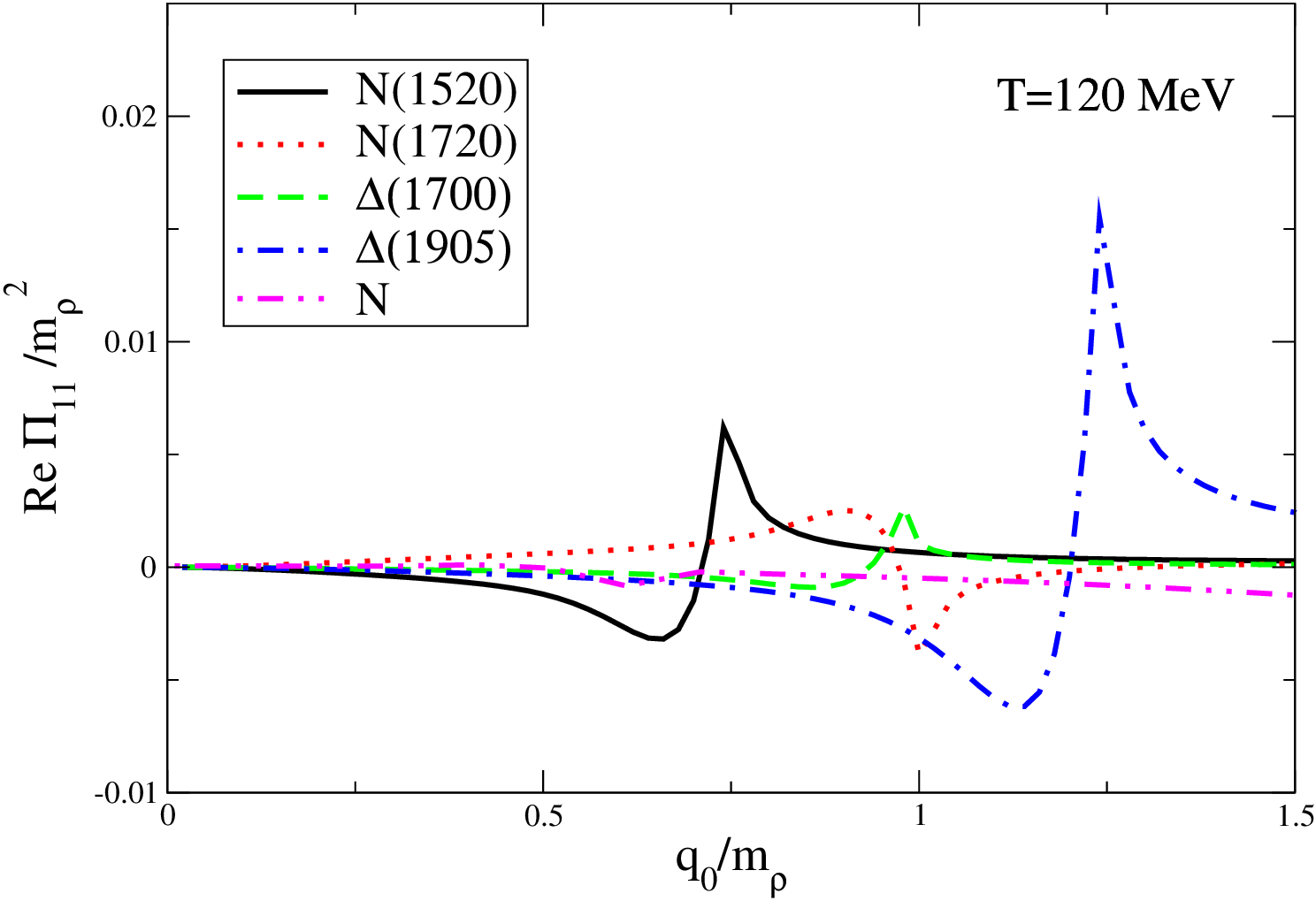}}
\par}
\caption{Contribution to the $\rho$ self-energy from scattering with
  nucleons and various other nucleon resonances. The main contribution
  for $q_0\sim m_\rho$ stems  from the formation of $s$-channel,
  positive parity spin 3/2 N(1720) and spin 5/2 $\Delta$(1905)
  resonances.} 
\label{fig1}
\end{figure}
%%%%%%%%%%%%%%%%%%%%%%%%%%%%%%%%%%%%%%%%%%%%%%%%%%%%%%%%

A look at the review of particle physics~\cite{Hagiwara} reveals the
existence of four baryon resonances with rest masses near the sum of
the rest masses of $\rho$ and nucleon (N) and with sizable decay rates
into the $\rho$--N channel. These are N(1520), N(1720),
$\Delta$(1700) and $\Delta$(1905). Table~I shows their quantum numbers
and branching ratios into the $\rho$--N channel as well as the
values of the coupling constants used in the calculation. These last
are computed by adjusting the experimentally measured branching ratios
into the $\rho$--N channel to the theoretical expression for the decay
width obtained by using the corresponding interaction Lagrangian. For
a reliable estimate of the coupling constants, we include the finite
width of $\rho$ by folding the expression for the width at a given
$\rho$ mass with the $\rho$ spectral function. For
definitiveness, the $\rho$ spectral function is taken as a relativistic
Breit--Wigner function
\be
   S(q)=\frac{2m_\rho\Gamma (q)}{(q^2-m_\rho^2)^2+ (m_\rho\Gamma(q)
   )^2}\, , 
   \label{BW}
\ee
where we also include the proper phase space angular momentum
dependence for the $\rho$ decay into two
pions~\cite{Peters}, taking  
\be
   \Gamma (q)=\Gamma^{\mbox{\tiny{vac}}}\left(
   \frac{\sqrt{q^2/4-m_\pi^2}}{\sqrt{m_\rho^2/4-m_\pi^2}}\right)^3\, ,
\ee
where we use $\Gamma^{\mbox{\tiny{vac}}}=150$ MeV.
For comparison, Table~I also shows the values of the 
coupling constants $f_{J^P}^{\mbox{\tiny{N--R}}}$ obtained by means of
a non-relativistic approach~\cite{Peters}. The interaction Lagrangians
${\mathcal {L}}$ are given by
\be
   {\mathcal {L}} = \left\{ 
   \begin{array}{lr}
      \frac{f_{3/2^-}}
      {m_\rho}{\mathcal {F}}_{\rho N3/2^-}
      \bar{\psi}^\mu\gamma^\nu\psi F_{\mu\nu} &
      \left(\ J^P=\frac{3}{2}^-\right) \\
      \frac{f_{3/2^+}}
      {m_\rho}{\mathcal {F}}_{\rho N3/2^+}
      \bar{\psi}^\mu\gamma^5\gamma^\nu\psi F_{\mu\nu} &
      \left(\ J^P=\frac{3}{2}^+\right) \\
      \frac{f_{5/2^+}}
      {m_\rho^2}{\mathcal {F}}_{\rho N5/2^+}
      \bar{\psi}^{\mu\nu}\gamma^5\gamma^\lambda
      \psi\partial_\nu F_{\mu\lambda} &
      \left(\ J^P=\frac{5}{2}^+\right) \\ 
   \end{array}
   \right.
   \label{LagBar}
\ee
where $f_{J^P}$ are the coupling constants between
$\rho$, N and the baryon resonance $R$,
$F_{\mu\nu}=\partial_\mu\rho_\nu - \partial_\nu\rho_\mu$ is the $\rho$
field strength tensor, $\psi$ is the nucleon field, $\psi^\mu$ is the
spin 3/2 field and $\psi^{\mu\nu}$ is the spin 5/2
field~\cite{Teodorescu, Rushbrooke}. ${\mathcal {F}}_{\rho NR}$ are
hadronic form factors that take into account the finite size of the
particles that appear in the effective vertexes. These form factors
are taken to be of dipole form
\be
   {\mathcal {F}}_{\rho NR}=\left(\frac{2\Lambda^2 + m^2_R}{2\Lambda^2
   + s}\right)^2\, ,
   \label{dipole}
\ee
where $s$ is the energy squared in the system where the resonance $R$
is at rest and $\Lambda$ is a phenomenological cutoff. The obtained
values for the coupling constants do not change when $\Lambda$ varies
in the range $1$ GeV $<\Lambda<$ $2$ GeV which is a reasonable
interval when considering hadronic processes. All the interactions in
Eqs.~(\ref{LagBar}) are current and parity conserving. 

To compute the one-loop $\rho$ self-energy we use the 
generalized Rarita-Schwinger propagators for fields with spin higher
than 1/2, given by~\cite{Teodorescu, Brudnoy, Napsuciale}
\begin{widetext}
\be
   {\mathcal {R}}^{\mu\nu}_{3/2}(K)&=&\frac{(K\!\!\!\!\!\!\!
   \not\,\,\,\ +\ m_R)}{K^2 - m_R^2}
   \left\{-g^{\mu\nu} + \frac{1}{3}\gamma^\mu\gamma^\nu
   + \frac{2}{3}\frac{K^\mu K^\nu}{m_R^2} - 
   \frac{1}{3}\frac{K^\mu\gamma^\nu - K^\nu\gamma^\mu}{m_R}
   \right\}\label{firsRS}\\
   {\mathcal {R}}^{\alpha\beta\rho\sigma}_{5/2}(K)&=&
   \sum_{\stackrel{\textstyle{\rho\leftrightarrow\sigma}}
   {\textstyle{\alpha\leftrightarrow\beta}}}
   \left\{\frac{1}{10}\frac{K^\alpha K^\beta K^\rho K^\sigma}{m_R^4}
   + \frac{1}{10}\frac{K^\alpha K^\beta K^\sigma\gamma^\rho - 
   K^\rho K^\sigma K^\alpha\gamma^\beta}{m_R^3} 
   + \frac{1}{10}\frac{K^\alpha\gamma^\beta K^\sigma\gamma^\rho}
   {m_R^2}\right.\nonumber\\ 
   &+&
   \frac{1}{20}\frac{K^\alpha K^\beta g^{\sigma\rho} 
   + K^\sigma K^\rho g^{\alpha\beta}}{m_R^2} - \frac{2}{5}
   \frac{K^\alpha K^\sigma g^{\beta\rho}}{m_R^2}
   - \frac{1}{10}
   \frac{\gamma^\sigma K^\beta g^{\alpha\rho} - 
   \gamma^\alpha K^\sigma g^{\beta\rho}}{m_R}\nonumber\\
   &-&\left.\frac{1}{10}
   \gamma^\alpha\gamma^\sigma g^{\beta\rho} -
   \frac{1}{20}g^{\alpha\beta}g^{\sigma\rho} + 
   \frac{1}{4}g^{\alpha\rho}g^{\beta\sigma}\right\}
   \frac{(K\!\!\!\!\!\!\!
   \not\,\,\,\ +\ m_R)}{K^2 - m_R^2}
   \label{rarita}
\ee
\end{widetext}
where $m_R$ is the mass of
the resonance and the sum over the indexes $\alpha\beta\rho\sigma$ of
a tensor $T^{\alpha\beta\rho\sigma}$ means
\be
   \sum_{\stackrel{\textstyle{\rho\leftrightarrow\sigma}}
   {\textstyle{\alpha\leftrightarrow\beta}}} T^{\alpha\beta\rho\sigma}
   \equiv T^{\alpha\beta\rho\sigma} + T^{\beta\alpha\rho\sigma} 
   + T^{\alpha\beta\sigma\rho} + T^{\beta\alpha\sigma\rho}\, .
   \label{defT}
\ee
In order to ensure that the unphysical spin--1/2 degrees of freedom
contained in $\psi^\mu$ and $\psi^{\mu\nu}$ have no observable effects
even in the interacting theories described by Eqs.~(\ref{LagBar}),
the propagators in Eqs.~(\ref{firsRS}) and~(\ref{rarita}) have to be
regarded as the leading order terms in an expansion in the parameter
$1/m_B$ where $m_B$ is the (heavy) baryon mass~\cite{Napsuciale}.

We also include the contribution from interactions between $\rho$ and
nucleons given by
\be
   {\mathcal {L}}_{\mbox{\tiny{$\rho NN$}}}
   =f_{\mbox{\tiny{$\rho NN$}}}
   \bar{\psi}\left(\gamma^\mu -
   \frac{\kappa}{2m_N}\sigma^{\mu\nu}\partial_\nu\right)
   \rho_\mu\psi
   \label{lagrNN}
\ee
where $\sigma^{\mu\nu}=(i/2)[\gamma^\mu,\gamma^\nu]$, $m_N$ is the mass
of the nucleon and we take the values of the dimensionless coupling
constants $f_{\mbox{\tiny{$\rho NN$}}}$ and $\kappa$ 
as $f_{\mbox{\tiny{$\rho NN$}}}=2.63$ and $\kappa=6.1$~\cite{Machleidt}.

%%%%%%%%%%%%%%%%%%%%%%%%%%%%%%%%%%%%%%%%%%%%%%%%%%%%%%%%%%%%%%%%%%%
\begin{table}[b!]
\begin{tabular}{|ccccccc|}  \hline\hline
\multicolumn{1}{|c}{$R$}  &
\multicolumn{1}{c}{$J^P$} &
\multicolumn{1}{c}{$\rho h$ decay} &
\multicolumn{1}{c}{$\Gamma^{\mbox{\tiny{vac}}}_{\rho h}$} &
\multicolumn{1}{c}{$\Gamma^{\mbox{\tiny{vac}}}_{\mbox{\tiny{tot}}}$} &
\multicolumn{1}{c}{$g_{\rho hR}$} &
\multicolumn{1}{c|}{$IF$} \\ 
& & & ${\mbox{(MeV)}}$ & ${\mbox{(MeV)}}$ & ${\mbox{(GeV)}}^{-1}$ &
\\ \hline
$\omega$(782) & $1^-$ & $\rho\pi$ & $\sim 5$ & $8.43$ & 25.8 & 1 \\
$h_1$(1170)   & $1^+$ & $\rho\pi$ &   seen   & $\sim 360$ & 11.37
& 1 \\
$a_1$(1260)   & $1^+$ & $\rho\pi$ & dominant & $\sim 400$ & 13.27 
& 2 \\ 
$K_1$(1270)   & $1^+$ & $\rho K$  & $\sim 60$& $\sim 90$  & 9.42
& 2 \\
$\pi'$(1300)  & $0^-$ & $\rho\pi$ &   seen   & $\sim 400$ & 7.44
& 2 \\ \hline
\end{tabular}
\label{tab2}
\caption{List of meson resonances included in the calculation. The
  last column corresponds to the isospin factor accounting for the
  number of isospin channels that take part in the dispersion. The
  coupling constants are taken from the analysis in Ref.~\cite{Rapp}.}
\vspace{-0.5cm}
\end{table}
%%%%%%%%%%%%%%%%%%%%%%%%%%%%%%%%%%%%%%%%%%%%%%%%%%%%%%%%%%%%%%%%%%%%%%%

For the interaction Lagrangians in Eqs.~(\ref{LagBar}), the
calculation involves the sum two Feynman diagrams whose corresponding
expressions, written in Minkowski space are
\be
   \Pi^{\pm (a)}_{(J)\mu\nu} &=& IF\int\frac{d^4p}{(2\pi )^4}
   \frac{T^{\pm}_{(J)\mu\nu}(P,P+Q)}{[P^2-m_N^2][(P+Q)^2-m_R^2]}
   \nonumber\\
   \Pi^{\pm (b)}_{(J)\mu\nu} &=& IF\int\frac{d^4p}{(2\pi )^4}
   \frac{T^{\pm}_{(J)\mu\nu}(P,P-Q)}{[P^2-m_N^2][(P-Q)^2-m_R^2]}
   \label{loopint}
\ee
where $\pm$ refer to the case of interactions with
positive and negative parity baryon resonances, respectively, $IF$ is
the isospin factor and
\be
   \!\!\!\!\!\!\!\!T^{\pm}_{(3/2)\mu\nu}(P,K)&\!\!\!\!=\!\!\!\!&
   {\mbox{Tr}}[(P\!\!\!\!\!\!\!\not\,\,\,\,
   + m_N)\Gamma^{\pm}_{\mu\alpha}{\mathcal {R}}^{\alpha\beta}_{3/2}(K)
   \Gamma^{\pm}_{\beta\nu}]\nonumber\\
   \!\!\!\!\!\!\!\!T^{\pm}_{(5/2)\mu\nu}(P,K)&\!\!\!\!=\!\!\!\!&
   {\mbox{Tr}}[(P\!\!\!\!\!\!\!\not\,\,\,\,
   + m_N)\Gamma^{\pm}_{\mu\alpha\beta}
   {\mathcal {R}}^{\alpha\beta\rho\sigma}_{5/2}(K)
   \Gamma^{\pm}_{\rho\sigma\nu}]
   \label{tensors}
\ee
where the vertices $\Gamma^{\pm}_{\mu\alpha}$ and
$\Gamma^{\pm}_{\mu\alpha\beta}$, as obtained from the interaction
Lagrangians in Eqs.~(\ref{LagBar}), are given by
\be
   \Gamma^{\pm}_{\mu\alpha}&=&
   \left(\frac{f_{3/2^\pm}}{m_\rho}\right)
   {\mathcal {F}}_{\rho N3/2^\pm}\gamma_5^{(1\pm 1)/2}
   \left(\gamma_\mu Q_\alpha - Q\!\!\!\!\!\!\not\,\,\,\,g_{\mu\alpha}
   \right)\nonumber\\
   \Gamma^{\pm}_{\mu\alpha\beta}&=&
   \left(\frac{f_{5/2^\pm}}{m_\rho}\right)
   {\mathcal {F}}_{\rho N5/2^\pm}\gamma_5^{(1\pm 1)/2}\gamma_\delta Q^\beta
   \nonumber\\
   &&\left(Q_\alpha g_{\delta\mu} - Q_\delta g_{\alpha\mu}\right)\, .
   \label{vertices}
\ee

Figure~\ref{fig1} shows the temperature dependent real part of
$\Pi^{11}$ scaled to the square of the $\rho$ mass in vacuum, for each
of the resonances listed in Table~I as a function 
of $q_0/m_\rho$, where $q_0$ is the energy of the $\rho$ meson at
rest, for a temperature $T=120$ MeV. Notice that the two terms in
Eqs.~(\ref{loopint}) represent the contribution from nucleons and
anti-nucleons to the scattering, as corresponds to the scenario where
resonance production happens from the --almost baryon free-- central
region of the reaction. The isospin factor considered for all these
processes has been taken as $IF=2$. The main contributions in
magnitude for $q_0\sim  m_\rho$ come from the resonances with even
parity N(1720) and $\Delta$(1905). We also show the contribution from
scattering with nucleons, taking $m_N$ to its vacuum value. Notice
that for the kinematical range considered, the contribution from
nucleons is completely negligible. 

An interesting aspect of the result is the difference in the behavior
between the real parts of the contributions of N(1720) and $\Delta
(1905)$ to the $\rho$ self-energy as a function of $q_0$, the former
starting out repulsive and the latter attractive. The reason for this
behavior is that, given that these are resonances with different spin,
the structure of their propagators and 
couplings with nucleons is different. The leading term for each case when
$q_0 \rightarrow 0$ is of the form $c*q_0$ where $c$ is a numerical
coefficient to which several terms from the product of the propagator
and vertices contribute. It turns out that this coefficient is
positive in the case of N(1720) and negative in the case of $\Delta
(1905)$. We emphasize that this conclusion is born out of the
explicit calculation. We should however point out that an important
cross check of the result, namely, the transversality of the
self-energy, has been carried out. This is by no means a trivial check
of the consistency of the calculation since had one or more of the
terms that make up the above mentioned $c$ coefficient been wrong, the
transversality would have been spoiled.

\section{Interactions of $\rho$ with pions and mesons}\label{secIII}

We now look at the contribution to the $\rho$ self-energy stemming from
scattering with pions and other mesons. Table~II shows the
quantum numbers of those mesons with sizable branching ratios
involving the pion and $\rho$ and whose rest mass is near the sum of
the rest masses of the pion/kaon and $\rho$, as well as the coupling
constants used in the calculation. For the interaction of $\rho$ and
the pion, $\pi$, we take the Lagrangian
\be
   {\mathcal {L}} = g_{\mbox{\tiny{$\rho\pi\pi$}}}
                    \pi(\stackrel{\leftrightarrow}{\partial}\ \!\!\! ^\mu
                     - ig_{\mbox{\tiny{$\rho\pi\pi$}}}\rho^\mu)\pi
                    \rho_\mu
   \label{lagpipirho}
\ee
obtained by {\it gauging} the pion-pion interaction
Lagrangian~\cite{Gale, Ayala}. The value of the dimensionless coupling
constant, as determined by the $\rho$ width in vacuum, is taken as
$g_{\mbox{\tiny{$\rho\pi\pi$}}}=6.06$~\cite{Gale}. 

%%%%%%%%%%%%%%%%%%%%%%%%%%%%%%%%%%%%%%%%%%%%%%%%%%%%%%%%
\begin{figure}[t!] % fig2
\vspace{0.4cm}
{\centering
\resizebox*{0.4\textwidth}
{0.2\textheight}{\includegraphics{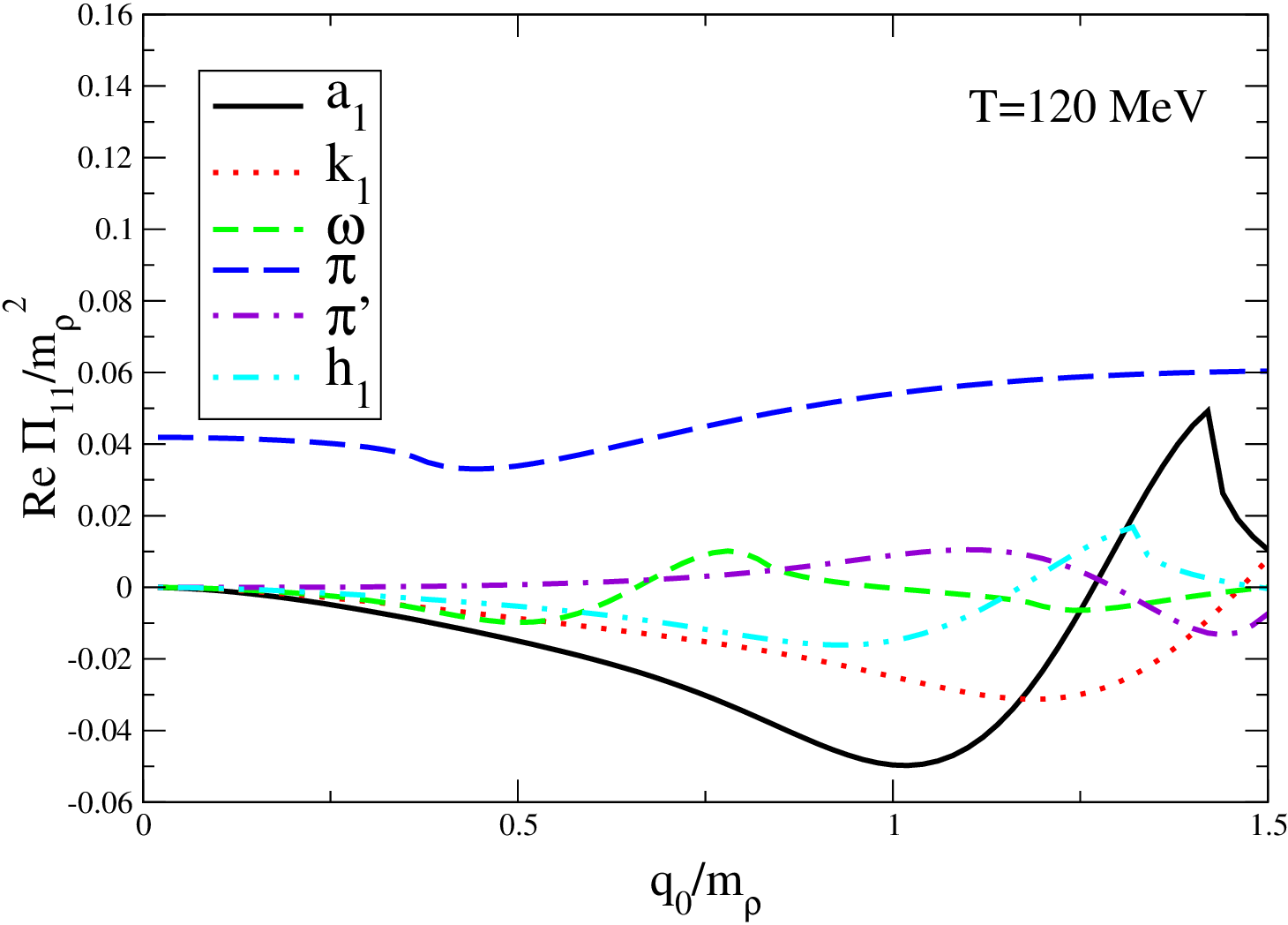}}
\par}
\caption{Contribution to the $\rho$ self-energy from scattering with
  pions and various other mesons. The main contribution for $q_0\sim
  m_\rho$ stems from the formation of an $s$-channel axial-vector
  resonance $a_1$.} 
\label{fig2}
\end{figure}
%%%%%%%%%%%%%%%%%%%%%%%%%%%%%%%%%%%%%%%%%%%%%%%%%%%%%%%%

The interaction Lagrangians involving the other relevant mesons are
taken as~\cite{Rapp}  
\be
   {\mathcal {L}} = \left\{ 
   \begin{array}{l}
   g_{\mbox{\tiny{$\rho PA$}}}{\mathcal {F}}_{\rho PA}
   \ A^\mu\partial^\nu P F_{\mu\nu} \\
   g_{\mbox{\tiny{$\rho\pi V$}}}{\mathcal {F}}_{\rho\pi V}
   \ \epsilon_{\mu\nu\sigma\tau}
   \partial^\mu \pi V^\nu\partial^\sigma\rho^\tau\\
   (g_{\mbox{\tiny{$\rho\pi P'$}}}/m_\rho){\mathcal {F}}_{\rho\pi P'} 
   \ \partial^\mu \pi\partial^\nu P'F_{\mu\nu}\\ 
   \end{array}
   \right.
   \label{lagothermesons}
\ee 
where $V^\mu$ and $A^\mu$ represent the vector and axial-vector fields
and $P$, $P'$ the pseudo-scalar fields. ${\mathcal {F}}_{\rho PR}$ 
are dipole form factors [see Eq.~(\ref{dipole})]. The coupling
constants $g_{\mbox{\tiny{$\rho PR$}}}$ in
Eqs.~(\ref{lagothermesons}) are taken from 
Ref.~\cite{Rapp}. The interaction Lagrangians in 
Eqs.~(\ref{lagpipirho}) and~(\ref{lagothermesons}) are current and
parity conserving as well as compatible with chiral symmetry.

The expressions for the one-loop $\rho$ self-energy corresponding
to the interaction Lagrangians in Eqs.~(\ref{lagothermesons}), can be
written in Minkowski space, as 
\be
   \Pi^{(J^P)}_{\mu\nu} &\!\!\!=\!\!\!& IF\int\frac{d^4p}{(2\pi )^4}
   \frac{M^{(J^P)}_{\mu\nu}}{[P^2-m_{\mbox{\tiny{P}}}^2][(Q-P)^2-m_R^2]}
   \label{loopintmesons}
\ee
where $m_{\mbox{\tiny{P}}}$ is the mass of the pseudoscalar ($\pi$ or
$K$) and $m_R$ is the mass of the vector, axial-vector or
$\pi'(1300)$ and $IF$ is the isospin factor. The numerators in
Eq.~(\ref{loopintmesons}) are given by
\be
   M^{(1^\pm)}_{\mu\nu}&=&\Gamma_{\mu\alpha}^{(1^\pm)}
   \left(g^{\alpha\beta} - \frac{K^\alpha K^\beta}{m_{R}^2}\right)
   \Gamma_{\beta\nu}^{(1^\pm)}\nonumber\\
   M^{(0^+)}_{\mu\nu}&=&\Gamma_\mu^{(0^+)}\Gamma_\nu^{(0^+)}\, ,
   \label{Mmesons}
\ee
where the vertices, as obtained from the interaction Lagrangians in
Eqs.~(\ref{lagothermesons}) are given by 
\be
   \Gamma_{\alpha\beta}^{(1^+)}&=&g_{\mbox{\tiny{$\rho PA$}}}
   {\mathcal{F}}_{\rho PA}\
   [g_{\alpha\beta}(P\cdot Q) - P_\alpha Q_\beta]\nonumber\\
   \Gamma_{\alpha\beta}^{(1^-)}&=&g_{\mbox{\tiny{$\rho\pi V$}}} 
   {\mathcal {F}}_{\rho\pi V}\ 
   [\epsilon_{\gamma\alpha\delta\beta}(Q-P)^\gamma Q^\delta]\nonumber\\
   \Gamma_{\alpha}^{(0^+)}&=&(g_{\mbox{\tiny{$\rho\pi P'$}}}/m_\rho) 
   {\mathcal {F}}_{\rho\pi P'}\nonumber\\
   &&[Q\cdot (Q-P)P_\alpha - (P\cdot Q)K_\alpha]
   \label{verticesmesons}
\ee

Figure~\ref{fig2} shows the temperature dependent real part of
$\Pi^{11}$ scaled to the square of the $\rho$ mass in vacuum arising
from pion exchange as well as each of the mesons listed in Table~II as
a function of $q_0/m_\rho$, for a temperature $T=120$
MeV. We notice that in the interval considered, the main contribution
comes from scattering of $\rho$ off pions. However, for $q_0\sim
m_\rho$ a sizable contribution in 
magnitude comes from $\pi$--$\rho$ scattering through the formation of
an $s$-channel axial-vector resonance $a_1$, which has the opposite
sign and about the same strength as the contribution from pion exchange, in
agreement with the findings in Ref.~\cite{Rapp}. Also, for $q_0\sim
m_\rho$, the rest of the contributions offset among themselves.

\section{Intrinsic mass of the $\rho^0$}\label{secIV}

\begin{figure}[t!] % fig3
\vspace{0.4cm}
{\centering
\resizebox*{0.4\textwidth}
{0.2\textheight}{\includegraphics{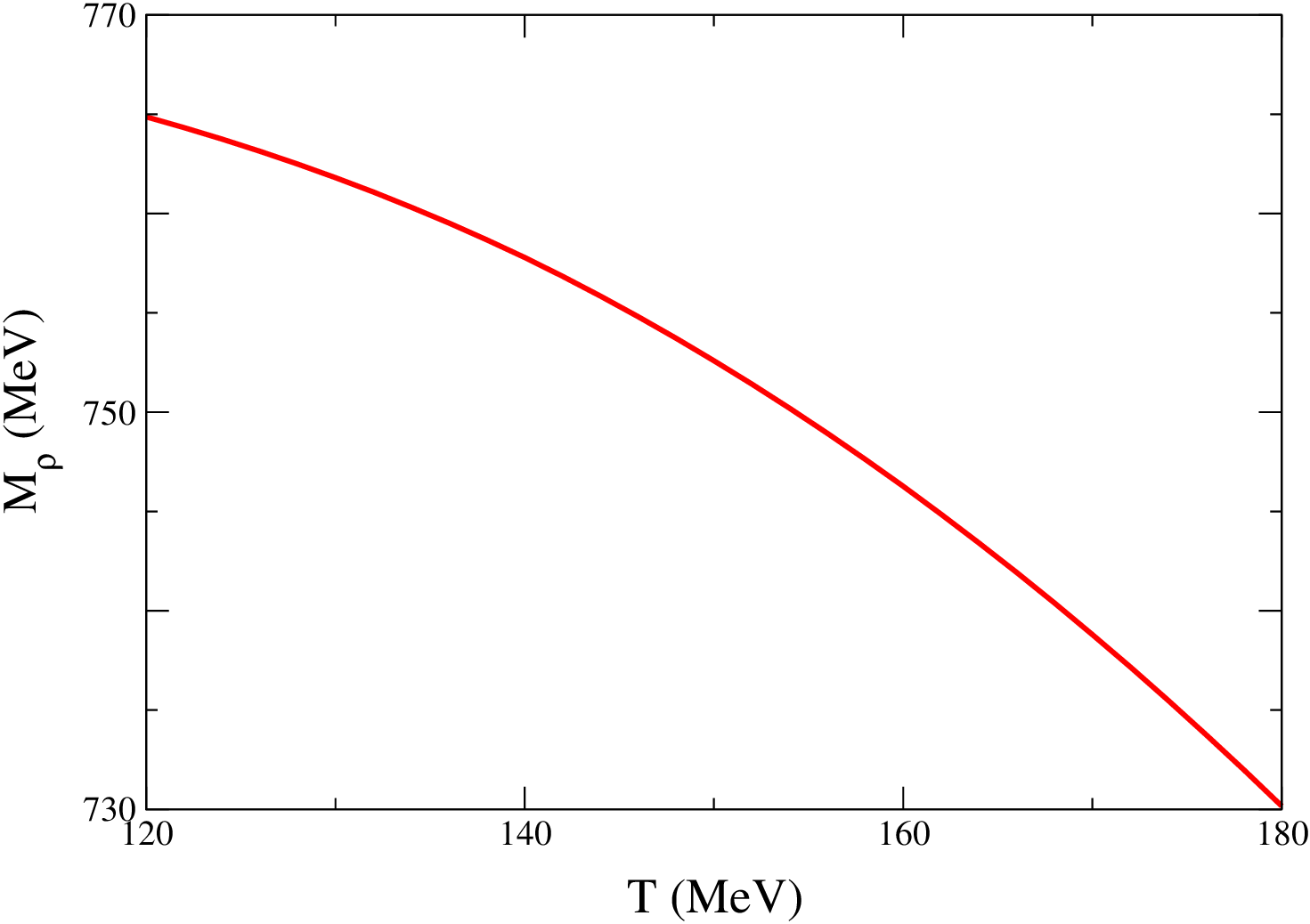}}
\par}
\caption{Intrinsic mass of $\rho^0$ as a function of temperature.}
\label{fig3}
\end{figure}

We now put together the contributions from all of the particles
considered in Secs.~II and~III. Figure~\ref{fig3} shows the total
shift in the intrinsic $\rho$ mass 
as a function of temperature. Notice that the shift increases in
magnitude as the temperature increases. For instance, taking
$m_\rho=770$ MeV, we get $M_\rho=764$--$730$ MeV when the temperature
varies between $T=120$--$180$ MeV, which is a reasonable range for the
temperature of the hadronic phase of a relativistic heavy-ion
collision between chemical and kinetic freeze-out.

%%%%%%%%%%%%%%%%%%%%%%%%%%%%%%%%%%%%%%%%%%%%%%%%%%%%%%%%
%%%%%%%%%%%%%%%%%%%%%%%%%%%%%%%%%%%%%%%%%%%%%%%%%%%%%%%%

We should emphasize that these findings refer to the {\em intrinsic}
changes in the $\rho$ mass. The overall change in the mass of the
peak of the invariant $\pi^+\ \pi^-$ distribution should contain
also the effects of phase-space distortions due to thermal motion of
the decay products as well as the effect due to the change in the
intrinsic $\rho$ width~\cite{Kolb, Rapp2}.  

\section{Summary and conclusions}\label{secV}

In this work we have computed the intrinsic changes in the $\rho$ mass
due to scattering with the relevant mesons and baryons in the context
of ultra-relativistic heavy-ion collisions, at finite temperature. We
have found that the contributions from scattering with nucleons through
the formation of even parity, spin 3/2 [N(1720)] and 5/2
[$\Delta$(1905)] nucleon resonances are significant. 

The different behavior between the real parts of the contributions of
N(1720) and $\Delta (1905)$ to the $\rho$ self-energy as a function of
$q_0$ is understood as arising from the different structure of their
propagators and couplings with nucleons, given that they are
resonances with different spin. 

These results underline the importance of scattering of $\rho$ mesons
with nucleons at finite temperature for the decrease of the intrinsic
mass of $\rho$, without the need of invoking a drop in the nucleon
mass during kinetic freeze-out. 

In conclusion, we have shown that it is possible to achieve a shift in
the intrinsic $\rho^0$ mass of up to $\sim - 40$ MeV, when
including the contributions of all the relevant mesons and baryons
that take part in the scattering, for temperatures within the commonly
accepted values between chemical and kinetic freeze-out.

\section*{Acknowledgments}

The authors are indebted to M. Kirchbach, R. Flores and M. Napsuciale
for useful comments and suggestions. A.A. also thanks R. Rapp for very
enlightening e-mail discussions. Support for this work has been
received in part by DGAPA-UNAM under PAPIIT grant number IN108001 and
CONACyT under grant number 40025-F and under a CONACyT-CNPq
bilateral agreement.

\end{document}